\documentclass[aps,prc,twocolumn,showpacs,superscriptaddress,footinbib]{revtex4-1}
\usepackage{graphicx}% Include figure files
\usepackage{dcolumn}% Align table columns on decimal point
\usepackage{bm}% bold math
\usepackage{enumerate}
\usepackage{amsmath}

\usepackage[normalem]{ulem}  % \sout{old text} for strikeout
\begin{document}

%\preprint{APS/123-QED}

\title{Isotopic dependence of the fragments' internal temperatures observed in multifragment emission}

\author{S.R. Souza}
\affiliation{Instituto de F\'\i sica, Universidade Federal do Rio de Janeiro Cidade Universit\'aria, 
\\CP 68528, 21941-972, Rio de Janeiro, Brazil}
\affiliation{Instituto de F\'\i sica, Universidade Federal da Bahia,
\\Campus Universit\'ario de Ondina, 40210-340, Salvador, Brazil}
\author{R. Donangelo}
\affiliation{Instituto de F\'\i sica, Universidade Federal do Rio de Janeiro Cidade Universit\'aria, 
\\CP 68528, 21941-972, Rio de Janeiro, Brazil}
\affiliation{Instituto de F\'\i sica, Facultad de Ingenier\'\i a, Universidad de la Rep\'ublica, 
Julio Herrera y Reissig 565, 11.300 Montevideo, Uruguay}

\date{\today}% It is always \today, today,
             %  but any date may be explicitly specified

\begin{abstract}
The internal temperatures of fragments produced by an excited nuclear source are investigated 
using the microcanonical version of the Statistical Multifragmentation Model, with discrete energy.
We focus on the fragments' properties at the breakup stage, before they have time to deexcite by 
particle emission.
Since the adopted model provides the excitation energy distribution of these primordial fragments, 
it allows one to calculate the temperatures of different isotope families and infer on the 
sensitivity to their isospin composition.
It is found that, due to the functional form of the nuclear density of states and the excitation 
energy distribution of the fragments, proton rich isotopes are hotter than neutron rich ones.
This property has been taken to be an indication of earlier emission of the former from a source 
that cools down as it expands and emits fragments.
Although this scenario is incompatible with the prompt breakup of a thermally 
equilibrated source, our results reveal that the latter framework also provides the same qualitative 
features just mentioned.
Therefore they suggest that this property cannot be taken as evidence for non-equilibrium emission.
We also found that this sensitivity to the isotopic composition of the fragments depends on the isospin 
composition of the source, and that it is weakened as the excitation energy of the source increases.
\end{abstract}

\pacs{25.70.Pq,24.60.-k}
                             % Classification Scheme.
%\keywords{Suggested keywords}%Use showkeys class option if keyword
                              %display desired
\maketitle

\begin{section}{Introduction}
\label{sect:introduction}
When nuclei collide at not too peripheral impact parameters, with incident energies per nucleon 
aobve a few tens of MeV, the overlap region reaches an excitation energy of several MeV per nucleon
and also has an appreciable increase of its nuclear density \cite{FlowSchussler,timeScaleDynStatistical2006,
exoticDens,XLargeSystems,AichelinPhysRep,Moretto1993,PawelFlow,BonaseraBUU,AMDReview2004}.
The succeeding fast expansion may lead the system to configurations of dynamical instabilities, during 
which many fragments are created \cite{timeScaleDynStatistical2006,exoticDens,XLargeSystems,AMDReview2004,
AichelinPhysRep,Moretto1993,PawelFlow,XLargeSystems,BorderiePhaseTransition2008,BonaseraBUU,
BurgioSpinodalInstabilities1994}.
Nevertheless, some dynamical calculations also suggest that the fragment composition is determined early 
in the expansion stage of the system \cite{mechanismsIMF1998,BondorfEarlyFragRecognition,
DorsoRandrupEarlyFragRecognition,DorsoAichelinEarlyFragRecognition}, before low densities have been attained.
In some studies \cite{XLargeSystems,mechanismsIMF1998},  it is also found that the properties of the system 
at the breakup stage assumed in the statistical models \cite{Bondorf1995,GrossMMM1990,BettyPhysRep2005,reviewSubal2001} 
are compatible with the configurations found in these dynamical calculations.
Although it gives support to a scenario in which the fragments are emitted statistically \cite{mechanismsIMF1998}, 
simulations based on the Antisymmetrized Molecular Dynamics Model \cite{AMDReview2004} suggest that a two stage model, 
in which fragments are formed in a prompt breakup and subsequently decay by particle emission, may not be 
the best representation of the actual process.
Indeed, it has been reported in Ref.\ \cite{timeScaleDynStatistical2006} that fragment deexcitation and 
fragment formation may take place concomitantly  during the process.
This scenario is, therefore, incompatible with the traditional hybrid treatments which separate the 
multifragment production in the two stages just mentioned \cite{smmWeisskopf2017}.

Efforts have been made to experimentally determine the properties of the fragments created in the breakup stage, 
right after the most violent stages of the collision \cite{freezeOut,primFragsIndra2003,primaryFrags2014,
primaryFrags2014_2,expRecEex2013,fragReconstructionNPA}.
Some of these experimental observations have been compared to Statistical Models \cite{Bondorf1995,MMM1997,MMM2002}, 
which adequately reproduced many experimental features \cite{expRecEex2013,freezeOut,primFragsIndra2003}.
However, other characteristics reported in different experimental studies, such as the saturation of the 
primary fragments' average excitation energies as a function of their atomic number, have not been 
satisfactorily accounted for by the statistical treatment employed in the analysis \cite{primFragsIndra2003}.
Similar conclusions may be drawn from the average excitation energies reported in Ref.\ \cite{expRecEex2013}, 
which yield the same results for different isotope families, as a function of their mass number.
The separation between the average excitation energies associated with different isotope families has not been 
reproduced by the statistical calculations, which predict a very weak isotopic dependence.  

In this work we examine the primary fragments' temperatures and focus on their isotopic dependence.
We employ the version of the Statistical Multifragmentation Model (SMM) presented in Ref.\ \cite{smmde2013}, 
which is built on the recurrence formulas developed in Ref.\ \cite{PrattDasGupta2000}, where the energy is 
treated as a discrete quantity.
In order to distinguish this version from the traditional SMM \cite{smm1,smm2,smm4}, we label it SMM-DE, 
emphasizing the discretization of the energy.
This SMM-DE is particularly useful to the present purpose as it provides the energy distribution of the primary 
fragments, rather than its average value as in the traditional SMM.
From such distributions and the fragments' density of states, one may calculate the average temperature of 
each species.
The present study is motivated by previous results \cite{expRecEex2013,CoolDyn2006,FriedmanLynch1983}, based 
on the Expanding Evaporating Source (EES) \cite{FriedmanLynch1983}, as well as from experimental studies 
\cite{expRecEex2013}, which suggest that proton rich isotopes are emitted earlier than neutron rich ones, 
so that the former are hotter than the latter.
We show that, although this conclusion is consistent with those results, other scenarios for the multifragment 
production, such as an equilibrium prompt breakup, also lead to the same observations.
Therefore, this aspect should not be considered as an evidence of non-equilibrium emission.

The remainder of the manuscript is organized as follows: The main fetures of the model are recalled in 
Sect.\ \ref{sect:model}.
The results are presented and discussed in Sect.\ \ref{sect:results}.
We conclude in Sect.\ \ref{sect:conclusions} with a brief summary.
\end{section}
 
\begin{section}{Theoretical framework}
\label{sect:model}
In Refs.\  \cite{ChaseMekjian1995,SubalMekjian,Subal1999}, efficient recurrence relations have been developed 
for the canonical ensemble which impose baryon number and charge conservation in each fragmentation mode.
Treating the system energy $E$ as a discrete quantity, in Ref.\  \cite{PrattDasGupta2000}, this formalism has 
been extended so that $E$ is also kept fixed in each fragmentation mode.
This allows its application to the microcanonical ensemble and an implementation based on the SMM has been 
developed in Ref.\ \cite{smmde2013}. 

In the framework of the SMM-DE, the total energy is then writen as $E=Q\Delta_Q$, where $Q$ is an integer 
number and $\Delta_Q$ is the granularity of the discretization.
The average fragment multiplicity, with mass and atomic numbers $a$ and $z$, respectively, and energy $q\Delta_Q$, , 
is given by \cite{PrattDasGupta2000}

\begin{equation}
\overline{n}_{a,z,q}=\frac{\omega_{a,z,q}}{\Omega_{A_0,Z_0,Q}}\Omega_{A_0-a,Z_0-z,Q-q}\;,
\label{eq:nazq}
\end{equation}

\noindent
where $A_0$ and $Z_0$ respectively represent the mass and atomic numbers of the decaying source.
The quantity $\Omega_{A,Z,Q}$ represents the number of states corresponding to the breakup of a nucleus 
$(A,Z)$ with energy $Q\Delta_Q$.
In Ref.\ \cite{PrattDasGupta2000}, it is shown that $\Omega_{A,Z,Q}$ can be calculated through the following 
recurrence relation:

\begin{equation}
\Omega_{A,Z,Q}=\sum_{\alpha,q_\alpha}\frac{a_\alpha}{A}\omega_{a_\alpha,z_\alpha,q_{\alpha}}\Omega_{A-a_\alpha,Z-z_\alpha,Q-q_{\alpha}}\;.
\label{eq:oazq}
\end{equation}

\noindent
The number of states of a nucleus $(A,Z)$ with energy $Q\Delta_Q$ is calculated through

\begin{equation}
\omega_{A,Z,Q}=\gamma_A\int_0^{\epsilon_{A,Z,Q}}\,dK\;\sqrt{K}\rho(\epsilon_{A,Z,Q}-K)\;,
\label{eq:wazq}
\end{equation}

\noindent
where

\begin{equation}
\gamma_A=\Delta_Q \frac{V_f (2m_n A)^{3/2}}{4\pi^2\hbar^3}\;,
\label{eq:gamma}
\end{equation}

\noindent
$m_n$ denotes the nucleon mass, $V_f$ is the free volume, and $\epsilon_{A,Z,Q}$ represents the sum of the 
fragment's kinetic and excitation energies.
The density of states $\rho(\epsilon^*)$ is described in Refs.\ \cite{smmde2013,internalTemperatures2015} 
and is built in such a way that it reproduces the behavior of the stantard SMM's Helmholtz free energy at 
high temperatures \cite{ISMMlong} and describes the experimental $\rho(\epsilon^*)$ at low excitation 
energies \cite{levelDensityGilbertCameron1965}.

Thus, the above relations allow us to calculate the primary fragment distribution for the breakup of a 
source at a fixed excitation energy $E$.
We will not provide further details on the model formulation in this work and, instead, refer the reader to 
Refs.\ \cite{smmde2013,internalTemperatures2015}, where a detailed presentation is made.

\end{section}

\begin{figure}[tb]
\includegraphics[width=9cm,angle=0]{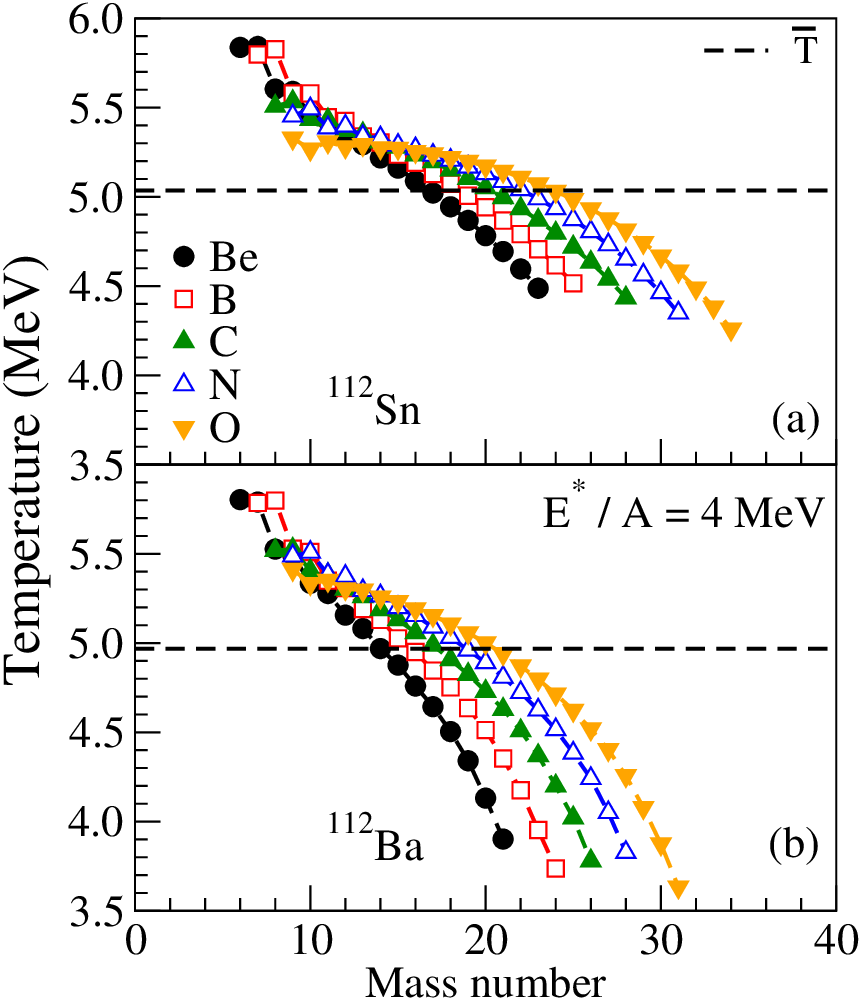}
\caption{\label{fig:tisoa} (Color online) Average temperature of light primary isotopes calculated with 
the SMM-DE for the (a) $^{112}$Sn and (b) $^{112}$Ba nuclei. 
The dashed line corresponds to the average breakup temperature of the source. 
For details see the text.}
\end{figure}

\begin{section}{Results}
\label{sect:results}
The breakup of the $^{112}$Sn and $^{112}$Ba nuclei at density equal to $\rho_0/3$, where $\rho_0$ 
corresponds to its saturation value, is discussed below.
We focus on the temperatures of the primary isotopes, which can be calculated through the standard 
thermodynamical relation

\begin{equation}
\frac{1}{T_{a,z,q}}=\frac{\partial\ln[\rho(\epsilon^*)]}{\partial\epsilon^*}\mid_{\epsilon^*=
\overline{\epsilon}^*_{a,z,q}}\;,
\label{eq:tint}
\end{equation}

\noindent
from which the average value is readily obtained with the help of Eq.\ (\ref{eq:nazq})

\begin{equation}
\overline{T}_{a,z}=\frac{\sum_q \overline{n}_{a,z,q}T_{a,z,q}}{\sum_q\overline{n}_{a,z,q}}\;.
\label{eq:aveTint}
\end{equation}

\noindent
The average excitation energy $\overline{\epsilon}^*_{a,z,q}$ is calculated through \cite{internalTemperatures2015}

\begin{equation}
\overline{\epsilon}^*_{a,z,q}=\frac{\gamma_a}{\omega_{a,z,q}}\int_0^{\epsilon_{a,z,q}}\,dK\;
(\epsilon_{a,z,q}-K)\sqrt{K}\rho(\epsilon_{a,z,q}-K)\;.
\label{eq:eex}
\end{equation}

The average temperatures of different primary isotopes predicted by the SMM-DE as a function of 
their atomic numbers are shown in Fig.\ \ref{fig:tisoa}(a) for the breakup of a $^{112}$Sn 
source with excitation energy $E^*/A = 4$ MeV.
The results exhibit a clear $A$ dependence, which weakens as $Z$ increases, and the proton rich 
isotopes are hotter than the neutron rich ones.
The average breakup temperature of the source is depicted in this figure by the dashed horizontal 
line and is obtained from

\begin{equation}
\frac{1}{T}=\frac{\partial\ln(\Omega_{A_0,Z_0,Q})}{\partial (Q\Delta_Q)}\approx 
\frac{\ln(\Omega_{A_0,Z_0,Q})-\ln(\Omega_{A_0,Z_0,Q-1})}{\Delta_Q}\;.
\label{eq:tbk}
\end{equation}

\noindent
One sees that important deviations from the average breakup temperature occur for neutron-deficient 
and neutron-rich isotopes.
In order to examine the sensitivity of the isospin composition of the source, Fig.\ \ref{fig:tisoa}(b) 
also displays $\overline{T}_{a,z}$ for isotopes produced in the breakup of the $^{112}$Ba nucleus at 
the same excitation energy per nucleon and density as the $^{112}$Sn nucleus.
The same features are once more observed but the magnitude of the effects is amplified.

\begin{figure}[tb]
\includegraphics[width=9cm,angle=0]{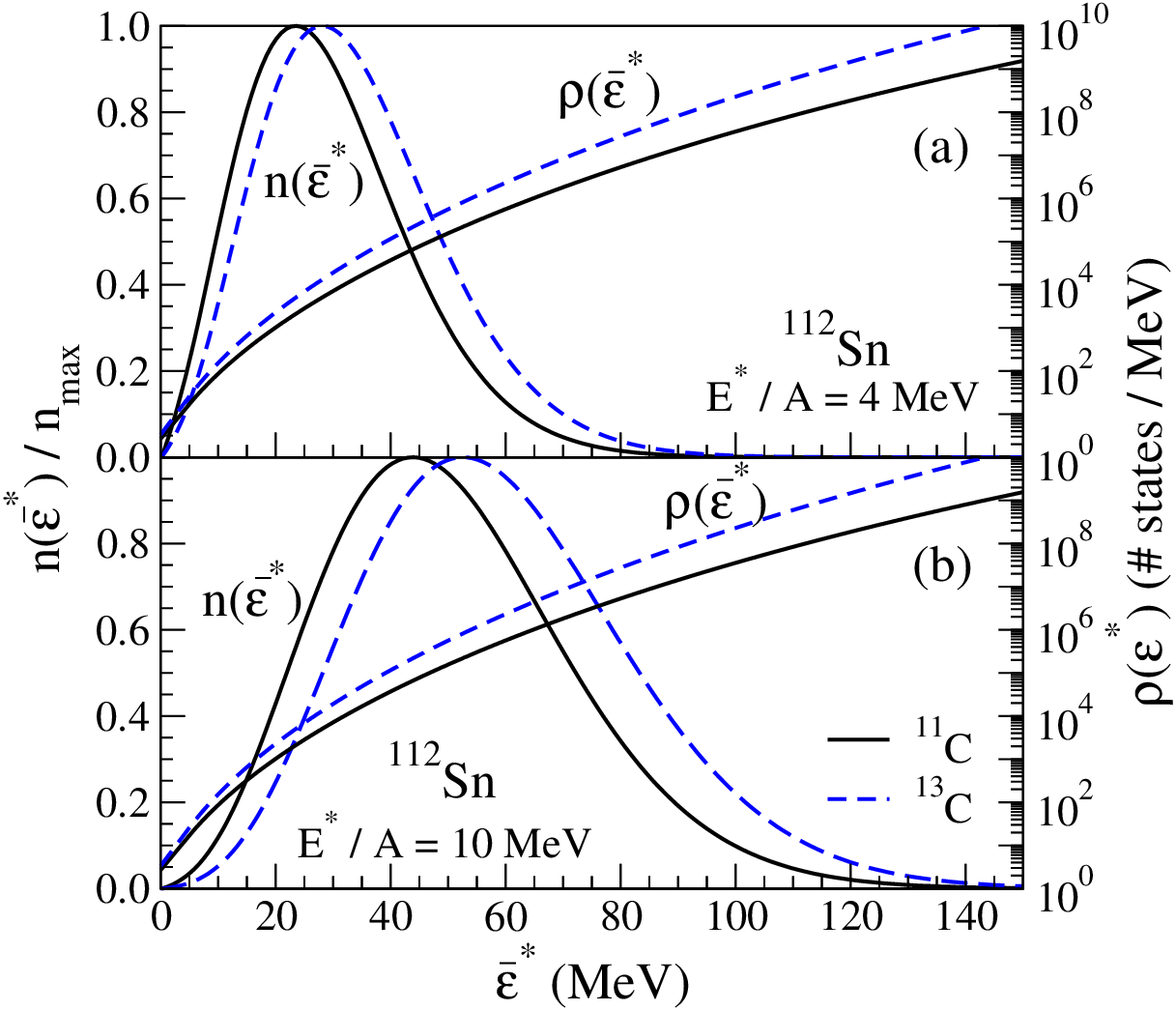}
\caption{\label{fig:eexrho} (Color online) Excitation energy distribution (left scale) and density 
of states (right scale) of the $^{11}$C and $^{13}$C isotopes produced in the breakup of the $^{112}$Sn 
nucleus at (a) $E^*/A = 4$ MeV and (b) $E^*/A = 10$ MeV. For details see the text.}
\end{figure}

To understand the behavior of the average temperatures just presented, we show, in Fig.\ \ref{fig:eexrho}(a), 
the  distribution of the average excitation energy of $^{11}$C and $^{13}$C isotopes produced 
in the breakup of a $^{112}$Sn nucleus at $E^*/A = 4$ MeV.
Both distributions are qualitatively similar, exhibiting a bell shape with the peak of the heavier 
isotope occurring at a slightly higher excitation energy value than the one associated with the lighter one.
The density of states $\rho(\overline\epsilon^*)$ for both fragments are also shown in this figure.
One sees that it exhibits a larger slope in the case of the $^{13}$C isotope in the region where 
the excitation energy distribution is non-negligible.
It thus leads to a smaller temperature value than in the case of the $^{11}$C isotope.
This explains the isotopic dependence of the fragments' temperatures observed in Fig.\ \ref{fig:tisoa}.

As the excitation energy of the source increases, the fragments' energy distributions is expected to
broaden and shift  towards higher excitation energy values.
This is indeed observed in Fig.\ \ref{fig:eexrho}(b) where the energy disrributions are shown for $E^*/A=10$ MeV.
Since the difference between the slopes of the density of states of these carbon isotopes becomes 
smaller as the energy increases, one should expect the mass dependence of the isotope temperature 
to weaken as the excitation energy of the source increases.
This is indeed observed in Fig.\ \ref{fig:tc}, which exhibits the ratio between the temperatures of 
different carbon isotopes to that of the $^8$C for the source's excitation energy $E^*/A=4$ and 10 MeV.
Thus we expect that the isotopic dependence of the temperature reported in this work should become 
negligible as the excitation energy of the source becomes large.

It is important to emphasize that, in the framework of the SMM-DE, all the primary fragments are 
produced simultaneously.
The decrease of the temperature of a given isotope with the increase of its neutron number is thus
explained, in the framework of this model, through the behavior of the fragments' excitation energy 
distributions and their densities of states.
As a consequence, our results suggest that this feature is not a signature of non-equilibrium process.

\begin{figure}[tb]
\includegraphics[width=9cm,angle=0]{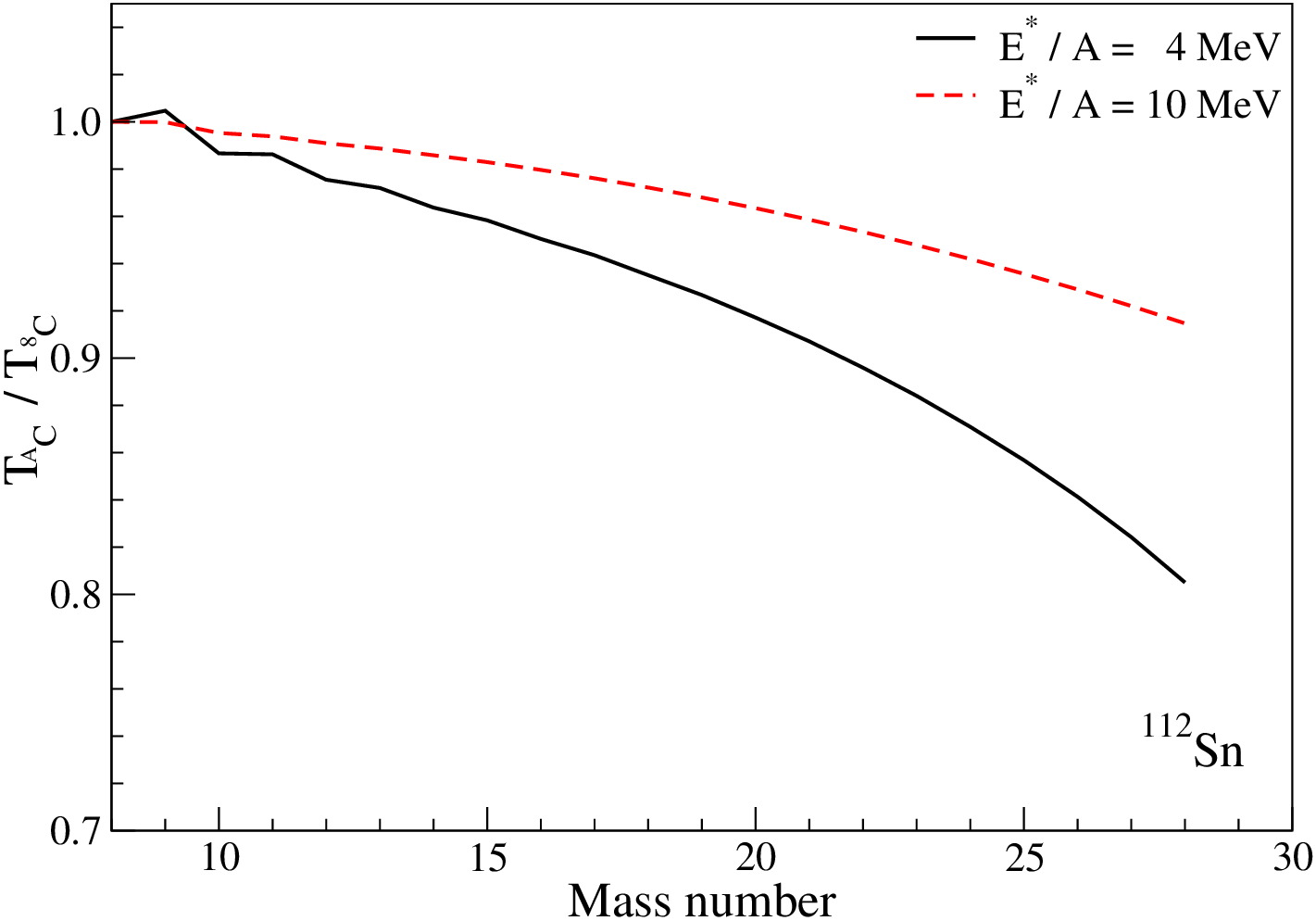}
\caption{\label{fig:tc} (Color online) Ratio between the temperatures of the Carbon isotopes to that 
of the $^8$C isotope.
The source is the $^{112}$Sn at $E^*/A = 4$ and 10 MeV. For details see the text.}
\end{figure}

\end{section}

\begin{section}{Concluding Remarks}
\label{sect:conclusions}
We have examined the isotopic dependence of the fragments' temperature in the framework of the SMM-DE.
This version of the model furnishes the excitation energy distribution, rather than the average value, 
of each species.
It thus allows one to calculate the internal temperatures of the primary fragments.
We found a fairly strong mass dependence of the temperature within each isotope family,
neutron poor isotopes being hotter than neutron rich ones.
This characteristic has been previously considered to be an indication of the existence of non-equilibrium 
effects, in particular that the proton rich isotopes were emitted earlier than the neutron rich ones.
Since, in the framework of the SMM, all of the primary fragments are produced simultaneously by a 
thermally equilibrated source, our results offer an alternative interpretation to this effect.
We also found that this sensitivity of the temperature to the isospin composition of the fragments 
becomes weaker as the excitation energy of the source increases, a prediction we consider could
be interesting to verify experimentally. 
 
\end{section}

\begin{acknowledgments}
This work was supported in part by the Brazilian
agencies Conselho Nacional de Desenvolvimento Científico
e Tecnológico (CNPq) and Fundação Carlos Chagas Filho de
Amparo à Pesquisa do Estado do Rio de Janeiro (FAPERJ),
a BBP grant from the latter. We also thank the
Programa de Desarrollo de las Ciencias Básicas (PEDECIBA)
and the Agencia Nacional de Investigaci\'on e Innovaci\'on
(ANII) for partial financial support.
\end{acknowledgments}

\bibliography{manuscript}
\bibliographystyle{apsrev4-1}

\end{document}